# Radiative heat transfer between metallic nanoparticle clusters in both near field and far field

Minggang Luo[1], Jian Dong[2], Junming Zhao[1*], Linhua Liu[1,2], Mauro Antezza[3,4]

*1 School of Energy Science and Engineering, Harbin Institute of Technology, 92 West Street, Harbin 150001, China*

*2 School of Energy and Power Engineering, Shandong University, Qingdao 266237, China*

*3 Laboratoire Charles Coulomb (L2C) UMR 5221 CNRS-Universit éde Montpellier, F-34095 Montpellier, France*

*4 Institut Universitaire de France, 1 rue Descartes, F-75231 Paris Cedex 05, France*

## ABSTRACT

Micro-nanoparticle systems have wide applications in thermal science and technology. In dense particulate system, the particle separation distance may be less than the characteristic thermal wavelength and near field effect will be significant and become a key factor to influence thermal radiation transfer in the system. In this study, radiative heat transfer (RHT) between two metallic nanoparticles clusters are explored using many-body radiative heat transfer theory implemented with the coupled electric and magnetic dipole (CEMD) approach, which effectively takes into account the contribution of magnetic polarization of metallic nanoparticles on heat exchange. As the focus, the effects of magnetic polarization and many-body interaction (MBI) on RHT were analyzed. The effects of fractal dimension and relative orientation of the clusters were also analyzed. Results show that the contribution of magnetically polarized eddy-current Joule dissipation dominates the RHT between Ag nanoparticle clusters. If only electric polarization (EP approach) is considered, the heat conductance will be underestimated as compared with the CEMD approach in both near field and far field regime. The effect of MBI on the RHT between Ag nanoparticle clusters is unobvious at room temperature, which is quite different from the SiC nanoparticle clusters. For the latter, MBI tends to suppress RHT significantly. The relative orientation has remarkable effect on radiative heat flux for clusters with lacy structure when the separation distance is in the near field. While for the separation distance in far field, both the relative orientation and the fractal dimension has a weak influence on radiative heat flux. This work will help the understanding of thermal transport in dense particulate system.

## I. INTRODUCTION

Due to rich physics and wide range of potential applications, particularly with the advancement of micro-nano technologies, near field radiative heat transfer (NFRHT) has received considerable attention in recent years. The fluctuational electrodynamics theory developed by Rytov et al. [1] was well recognized as a theoretical framework to predict NFRHT [2-8], which has been verified by many recent experimental observations [9-17]. In dense particulate system, the particle separation distance may be comparable to or less than the characteristic thermal wavelength, hence near field effect will be significant and become the key factor to influence the thermal radiation transfer characteristics.

Early studies on NFRHT mostly considered system consisting of two bodies, e.g., two plates, two particles, etc. Domingues et al. [18] investigated radiative thermal conductance in near field by means of molecular dynamics coupled with fluctuation dissipation theorem. Narayanaswamy et al. [19] studied the NFRHT between two spherical particles of arbitrary radius based on rigorous solution of the fluctuational electrodynamics theory with an quasi-analytical approach using vector spherical harmonics expansion. Czapla et al. [20] extended the method developed by Narayanaswamy et al. [19] to investigate NFRHT between two coated spheres with an arbitrary numbers of coatings. Messina et al. [21-23] proposed a scattering operator method to investigate NFRHT between two particles of arbitrary shape. Under dipole approximation, Chapuis et al. [24] took into consideration the contribution of magnetic-magnetic polarized eddy-current Joule dissipation (MM contribution) when investigating RHT between two particles. They showed that the EE contribution dominates the RHT between dielectric particles and the MM contribution dominates the RHT between metallic particles. Manjavacas et al. [25] considered the contribution of electromagnetic cross-terms, e.g. magnetic-electric polarized eddy-current Joule dissipation contribution (ME contribution) and electric-magnetic polarized displacement current dissipation (EM contribution) in calculating radiative heat flux between two spherical particles. For dimers consisting of two dielectric particles or two metallic particles, their research results were consistent with the work by Chapuis et al. [24].

For NFRHT in system consisting of many particles, some important progresses were reported only recently. There are very complex near field mutual interactions among particles and the approach to deal with NFNHT in two-body system can not be directly applied to the system of many particles. Ben-Abdallah et al. [26] developed a many-body radiative heat transfer theory to investigate RHT in many particles system and the effect MBI on RHT. Though the theory is based on dipole approximation, this approach is very general and can be effectively applied to predict NFRHT in a system of small particles of any shape, which allows detailed analysis of MBI in particulate system. They showed that radiative heat flux between two SiC particles can be enhanced significantly due to MBI after the insertion of a third particle [26]. The heat super-diffusion behavior induced by MBI in networks of spherical particles was also predicted [27]. It was also demonstrated that the spatial distribution of particles in a system of particles plays a key role in determining radiative heat flux [28]. In contrast to the enhancement effect of MBI on radiative heat flux observed in the system of three SiC particles, it was also reported that MBI inhibits the radiative heat flux in dielectric clusters of many particles [29].

Recently, there were some notable theoretical development to deal with NFRHT in system of particles. Krüger et al. [30] proposed a trace formulas and applied it to investigate RHT in many particles system composed of particles with arbitrary shape and radius. Müller et al. [31] extended the trace formulas to the many particles system embedded in a non-absorbing medium. Zhu et al. [32] investigated RHT in many particles system without the constraint of reciprocity by means of the trace formulas. Czapla et al. [33] derived formulas for NFRHT in a chain of spheres of arbitrary size, spacing, and isotropic optical properties based on the theoretical frame developed by Narayanaswamy et al. [19], which was validated by the thermal discrete dipole approximation (T-DDA) [34] and fluctuating surface currents (FSC)/boundary element methods (BEM) [35]. Becerril et al. [36] investigated near field energy transfer between three nanoparticles system modulated by coupled multipolar modes and found that coupled modes between nanoparticles provide more channels for NFRHT. By noticing the many-body radiative heat transfer theories did not include the mutual interactions of the electric and magnetic dipole moments and most of the studies considered dielectric particles with magnetic dipole moment neglected, Dong et al. [37] developed a coupled electric and magnetic dipole (CEMD) approach for the RHT in a collection of objects in mutual interaction, as an extension of the work of Ben-Abdallah et al. [26]. The CEMD approach takes all the EE, EM, ME and MM contributions to RHT into consideration, allows the analysis of NFRHT and the effect of MBI in system containing groups of metallic particles, where the magnetic terms may play an important role. Chen et al. [38] applied the CEMD approach to investigate RHT between two assembled systems of core-shell nanoparticles and observed similar inhibitive effect of MBI on total radiative heat flux as reported for dielectric particles by Dong et al. [29]. Previous studies have shown that the effect of MBI on RHT is complex in system of particles and significantly influences the radiative heat flux. It remains unclear about the effect of MBI on the RHT characteristics in system of metallic particles.

In this work, the RHT between two metallic nanoparticles clusters are explored using many-body radiative heat transfer theory with the CEMD approach, which effectively takes into account the contribution of magnetic response of metallic nanoparticles on heat exchange. The effect of magnetic polarization and many-body interaction on NFRHT in dense particulate system are analyzed as the focus. The effects of fractal dimension and relative orientation of the clusters on NFRHT are also analyzed. This work is organized as follows. In Section II, physical model of the fractal cluster and theoretical aspects of the CEMD approach are presented. The formulas to evaluate the effect of MBI on radiative heat exchange in two nanoparticles clusters are given. In Section III, the mechanism of RHT between metallic nanoparticle clusters, the effects of MBI, fractal dimension and relative orientation of clusters on RHT is analyzed.

## II. MODEL AND METHOD

### A. Nanoparticle cluster generation

RHT between two metallic nanoparticles clusters is considered. The nanoparticle cluster is described by the following typical statistical rule [39]

$$N_S = k_0 \left( \frac{R_g}{a} \right)^{D_f} \tag{1}$$

where $N_S$ is the number of monomers in the cluster, $D_f$ is the fractal dimension, $k_0$ is the prefactor, $a$ is the radius of the monomers and $R_g$ is the radius of gyration. The $D_f$ is the main factor that describes the compactness of the aggregate. Clusters with three different $D_f$ (1.8, 2.3 and 2.8) are generated by the open source program provided by Skorupski et al.[40], shown in Fig. 1(a)-(c). The number of realizations of clusters has been checked. For more details about the cluster generation, please refer to the previous work [29].The number of monomers in the aggregate ($N_S$) is set as 400. RHT between two identical absorbing and emitting clusters, of which temperatures are fixed at $T$ and $T+\delta T$, is investigated. Separating gap ($d$) between clusters is defined as the distance between the bottom of the upper cluster and the top of the lower cluster, shown in Fig. 1(d). Both $d$ between clusters and separation distance between monomers inside cluster edge to edge are larger than $2a$, which makes dipole approximation valid [26, 37].

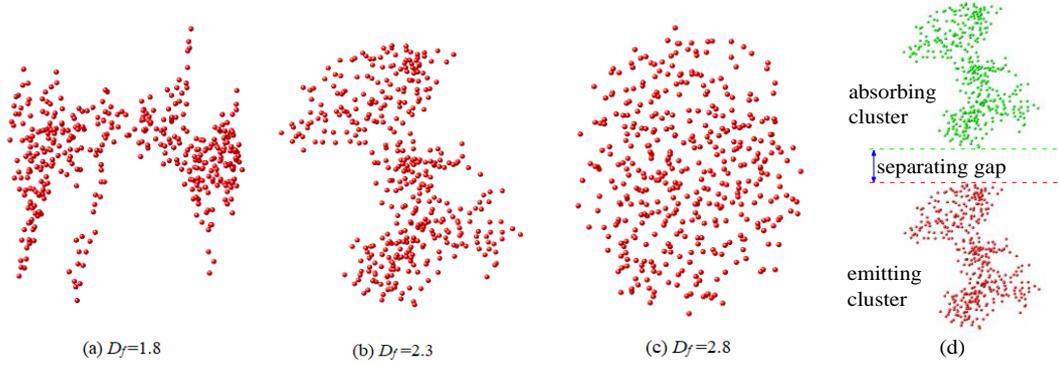

(a) $D_f$=1.8      (b) $D_f$=2.3      (c) $D_f$=2.8      (d)

**Fig. 1** Cluster configuration for three different fractal dimensions,
$N_S$ is 400 and radius of monomer is 5 nm. (a) $D_f$=1.8, (b) $D_f$=2.3, (c) $D_f$=2.8.
(d) schematic on definition of separating gap between the absorbing and emitting clusters.

## B. Polarizability of nanoparticle

In this work, Ag nanoparticle clusters are used for the calculations. SI unit system is used for all the formulation. For isotropic spherical particles, the electric dipole moment $\mathbf{p}$ and magnetic dipole moment $\mathbf{m}$ induced by the incident electromagnetic field in the vacuum read

$$\mathbf{p} = \varepsilon_0 \alpha_E \mathbf{E}^{inc} \tag{2}$$

$$\mathbf{m} = \alpha_H \mathbf{H}^{inc} \tag{3}$$

where $\alpha_E$ and $\alpha_H$ are electric and magnetic polarizability, $\mathbf{E}^{inc}$ and $\mathbf{H}^{inc}=\mathbf{B}^{inc}/\mu_0$ are the incident electric and magnetic fields, $\varepsilon_0$ is the vacuum dielectric permittivity. The electric and magnetic polarizabilities of a spherical Ag nanoparticle with radius of 5nm are shown in Fig. 2. Neglecting the high order absorption and scattering, $\alpha_E$ and $\alpha_H$ can

be obtained from the first order Lorentz-Mie scattering coefficients as

$$\alpha_E = \frac{i6\pi}{k^3} a_1 \tag{4}$$

$$\alpha_H = \frac{i6\pi}{k^3} b_1 \tag{5}$$

where $a_1$ and $b_1$ are the first order Lorentz-Mie scattering coefficients. The $n$th order Lorentz-Mie scattering coefficients are calculated from

$$a_n = \frac{\varepsilon j_n(y)[xj_n(x)]' - j_n(x)[yj_n(y)]'}{\varepsilon j_n(y)[xh_n^{(1)}(x)]' - h_n^{(1)}(x)[yj_n(y)]'} \tag{6}$$

$$b_n = \frac{j_n(y)[xj_n(x)]' - j_n(x)[yj_n(y)]'}{j_n(y)[xh_n^{(1)}(x)]' - h_n^{(1)}(x)[yj_n(y)]'} \tag{7}$$

where $x = kR$, $y = \sqrt{\varepsilon}kR$, $k$ is wave vector, $R$ is the particle radius, $\varepsilon$ is the dielectric permittivity, $j_n(x)$ and $h_n^{(1)}(x)$ are Bessel functions and the spherical Hankel functions. The dielectric permittivity of Ag is described by the Drude model [29, 41]

$$\varepsilon(\omega) = 1 - \frac{\omega_p^2}{\omega^2 + i\gamma\omega} \tag{8}$$

where $\omega$ is angular frequency, $\omega_p$ is $1.37\times10^{16}$rad s$^{-1}$ and $\gamma$ is $2.73\times10^{13}$rad s$^{-1}$. Note that the imaginary part of $\chi_E$ and $\chi_H$ are key factors in determining the exchanged radiative power according to the transmission coefficients defined by Eqs. (25) and (26). Localized surface plasmon resonance (LSPR) of metallic nanoparticle lies in the optical frequency, as shown in Fig. 2(a), which can't be excited thermally. Hence, RHT between metallic nanoparticles can't be as strong as the RHT between SiC dielectric particles, which can support low frequency localized surface phonon resonance (LSPhR). However, the magnetic response of the Ag nanoparticle is strong in the long wavelength range located near the thermal wavelength, indicating significant contribution of magnetic response to RHT in particulate system.

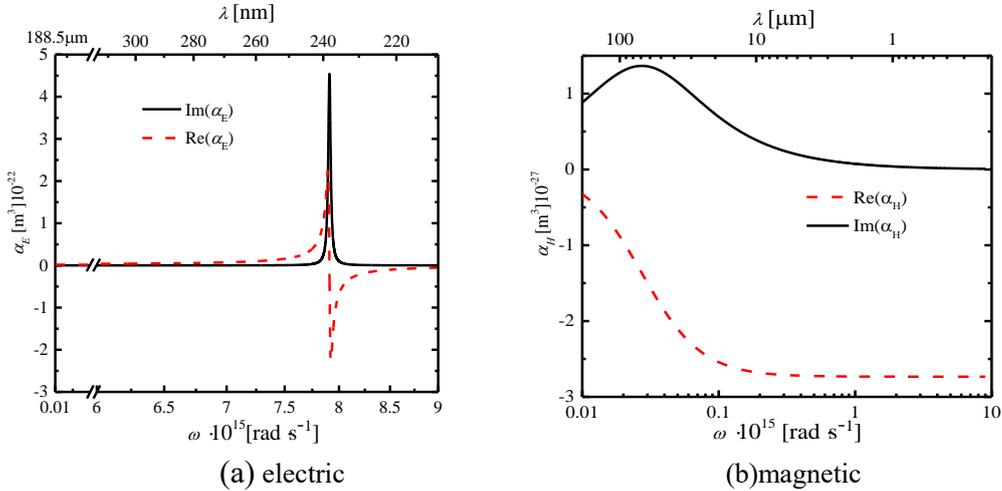

(a) electric          (b) magnetic

## C. Theoretical aspect

According to Poynting theorem, the power dissipation induced by the incident electromagnetic wave is [24]

$$P = \left\langle \frac{\partial \mathbf{p}}{\partial t} \cdot \mathbf{E}^{inc} - \mathbf{m} \cdot \frac{\partial \mathbf{B}^{inc}}{\partial t} \right\rangle \qquad (9)$$

where $<\ldots>$ means ensemble average, $P$ is power dissipation, $t$ is time, the first term in the right hand is electric polarized displacement current dissipation, the second term is the magnetic polarized eddy-current Joule dissipation. The corresponding cross-spectral density $P_\omega$ is

$$P_\omega = \omega \, \text{Im} \left\langle \mathbf{p} \cdot \mathbf{E}^{inc*} + \mathbf{m} \cdot \mathbf{B}^{inc*} \right\rangle \qquad (10)$$

where symbol * denotes the conjugation of the corresponding complex vector. For metallic nanoparticle, in addition to the electric contribution, the magnetic contribution to the power dissipation will be significant and even become dominant. In this work, the CEMD approach [37] is used to calculate RHT between metallic nanoparticle clusters, which effectively takes the EE, EM, ME and MM contributions to RHT into consideration and allows the analysis of NFRHT in system of metallic particles. In free space, the electromagnetic field at field point induced by an electric dipole $\mathbf{p}$ at source point are

$$\mathbf{E}_{EE} = \mu_0 \omega^2 G_0^{EE} \mathbf{p}, \quad \mathbf{H}_{ME} = k \omega G_0^{ME} \mathbf{p} \qquad (11)$$

where $\mu_0$ is the vacuum permeability, $G_0^{EE}$ and $G_0^{ME}$ are Green's function in free space.

$$G_0^{EE} = \frac{e^{ikr}}{4\pi r} [(1 + \frac{ikr - 1}{k^2 r^2}) \mathbb{I}_3 + \frac{3 - 3ikr - k^2 r^2}{k^2 r^2} \hat{\mathbf{r}} \otimes \hat{\mathbf{r}}] \qquad (12)$$

$$G_0^{ME} = \frac{e^{ikr}}{4\pi r} (1 - \frac{1}{ikr}) \begin{bmatrix} 0 & -\hat{r}_z & \hat{r}_y \\ \hat{r}_z & 0 & -\hat{r}_x \\ -\hat{r}_y & \hat{r}_x & 0 \end{bmatrix} \qquad (13)$$

where $\mathbb{I}_3$ is a 3×3 identity matrix, $r$ is magnitude of the separation vector $\mathbf{r} = \mathbf{r}_f - \mathbf{r}_s$ between the source point $\mathbf{r}_s$ and field point $\mathbf{r}_f$, $\hat{r}$ is the unit vector $\mathbf{r}/r$ and $\hat{r}_{v=x,y,z}$ denotes its three component, $\otimes$ denotes outer product of vectors. The electromagnetic field at $\mathbf{r}_f$ induced by the magnetic dipole $\mathbf{m}$ at $\mathbf{r}_s$ are

$$\mathbf{E}_{EM} = \mu_0 \omega k G_0^{EM} \mathbf{m}, \quad \mathbf{H}_{MM} = k^2 G_0^{MM} \mathbf{m} \qquad (14)$$

where $G_0^{EM} = -G_0^{ME}$ and $G_0^{MM} = G_0^{EE}$ are Green's function in free space. In many particles system, Green's function links the $j$th electromagnetic dipoles and their induced electromagnetic field at $i$th particle as

$$\mathbf{E}_{j\to i,EE} = \mu_0 \omega^2 G_{ij}^{EE} \mathbf{p}_j, \mathbf{H}_{j\to i,ME} = k\omega G_{ij}^{ME} \mathbf{p}_j \tag{15}$$

$$\mathbf{E}_{j\to i,EM} = \mu_0 \omega k G_{ij}^{EM} \mathbf{m}_j, \mathbf{H}_{j\to i,MM} = k^2 G_{ij}^{MM} \mathbf{m}_j \tag{16}$$

where $G_{ij}^{EE}$, $G_{ij}^{ME}$, $G_{ij}^{EM}$ and $G_{ij}^{MM}$ are the Green's functions in many particles system, which can be deduced from the Green's function in free space as follows.

$$\begin{pmatrix} 0 & \mathbb{G}_{12} & \cdots & \mathbb{G}_{1N} \\ \mathbb{G}_{21} & 0 & \ddots & \vdots \\ \vdots & \vdots & \ddots & \mathbb{G}_{(N-1)N} \\ \mathbb{G}_{N1} & \mathbb{G}_{N2} & \cdots & 0 \end{pmatrix} = \begin{pmatrix} 0 & \mathbb{G}_{0,12} & \cdots & \mathbb{G}_{0,1N} \\ \mathbb{G}_{0,21} & 0 & \ddots & \vdots \\ \vdots & \vdots & \ddots & \mathbb{G}_{0,(N-1)N} \\ \mathbb{G}_{0,N1} & \mathbb{G}_{0,N2} & \cdots & 0 \end{pmatrix} \boldsymbol{A}^{-1} \tag{17}$$

where elements in the matrix are given as

$$\mathbb{G}_{ij} = \begin{bmatrix} \mu_0 \omega^2 G_{ij}^{EE} & \mu_0 \omega k G_{ij}^{EM} \\ k\omega G_{ij}^{ME} & k^2 G_{ij}^{MM} \end{bmatrix}, \mathbb{G}_{0,ij} = \begin{bmatrix} \mu_0 \omega^2 G_{0,ij}^{EE} & \mu_0 \omega k G_{0,ij}^{EM} \\ k\omega G_{0,ij}^{ME} & k^2 G_{0,ij}^{MM} \end{bmatrix} \tag{18}$$

and $\boldsymbol{A}$ is a matrix including many-body interaction.

$$\boldsymbol{A} = \mathbb{I}_{6N} - \begin{bmatrix} 0 & \boldsymbol{\alpha}_1\mathbb{G}_{0,12} & \cdots & \boldsymbol{\alpha}_1\mathbb{G}_{0,1N} \\ \boldsymbol{\alpha}_2\mathbb{G}_{0,21} & 0 & \ddots & \vdots \\ \vdots & \vdots & \ddots & \boldsymbol{\alpha}_{N-1}\mathbb{G}_{0,(N-1)N} \\ \boldsymbol{\alpha}_N\mathbb{G}_{0,N1} & \cdots & \boldsymbol{\alpha}_N\mathbb{G}_{0,N(N-1)} & 0 \end{bmatrix}, \boldsymbol{\alpha}_i = \begin{bmatrix} \varepsilon_0\alpha_E^i\mathbb{I}_3 & 0 \\ 0 & \alpha_H^i\mathbb{I}_3 \end{bmatrix} \tag{19}$$

where $\boldsymbol{\alpha}_i$ is a 6×6 matrix, $\mathbb{I}_{6N}$ is a 6N×6N identity matrix. The power absorbed by $i$th particle excited with thermal emission from $j$th particle is

$$P_{j\to i} = 2\int_0^{+\infty} \frac{d\omega}{2\pi}\omega\left[\mathrm{Im}\left(\chi_E^i\right)\varepsilon_0\left\langle \mathbf{E}_{j\to i}^{inc}\cdot\mathbf{E}_{j\to i}^{inc*}\right\rangle + \mathrm{Im}\left(\chi_H^i\right)\mu_0\left\langle\mathbf{H}_{j\to i}^{inc}\cdot\mathbf{H}_{j\to i}^{inc*}\right\rangle\right] \tag{20}$$

where $\chi_E = \alpha_E - \dfrac{ik^3}{6\pi}\left|\alpha_E\right|^2$ and $\chi_H = \alpha_H - \dfrac{ik^3}{6\pi}\left|\alpha_H\right|^2$. The incident electromagnetic fields are excited by the thermal fluctuating electric dipole $\mathbf{p}_j^{fluc}$ and magnetic dipole $\mathbf{m}_j^{fluc}$.

$$\mathbf{E}_{j\to i}^{inc} = \mu_0\omega^2 G_{ij}^{EE}\mathbf{p}_j^{fluc} + \mu_0\omega k G_{ij}^{EM}\mathbf{m}_j^{fluc} \tag{21}$$

$$\mathbf{H}_{j\to i}^{inc} = k\omega G_{ij}^{ME}\mathbf{p}_j^{fluc} + k^2 G_{ij}^{MM}\mathbf{m}_j^{fluc} \tag{22}$$

With the application of the fluctuation dissipation theorem for electric and magnetic dipole moment [19, 34],

$$\left\langle p_{j,\alpha}^{fluc}\, p_{j,\beta}^{fluc*}\right\rangle = 2\frac{\varepsilon_0}{\omega}\operatorname{Im}(\chi_E^j)\Theta(\omega,T_j)\delta_{\alpha\beta} \tag{23}$$

$$\left\langle m_{j,\alpha}^{fluc}\, m_{j,\beta}^{fluc*}\right\rangle = \frac{2}{\omega\mu_0}\operatorname{Im}(\chi_H^j)\Theta(\omega,T_j)\delta_{\alpha\beta} \tag{24}$$

the power absorbed by the $i$th particle caused by $j$th particle can be written in Landauer-like formalism as

$$P_{j\to i} = 3\int_0^{+\infty}\frac{d\omega}{2\pi}\Theta(\omega,T_j)\mathcal{T}_{i,j}(\omega) \tag{25}$$

where $\mathcal{T}_{i,j}(\omega)$ is the transmission coefficient from the $j$th particle to the $i$th particle given as

$$\begin{aligned}
\mathcal{T}_{i,j}(\omega) = \frac{4}{3}k^4\Big[&\operatorname{Im}\left(\chi_E^i\right)\operatorname{Im}\left(\chi_E^j\right)\operatorname{Tr}\left(G_{ij}^{EE}G_{ij}^{EE\dagger}\right)\\
&+\operatorname{Im}\left(\chi_E^i\right)\operatorname{Im}\left(\chi_H^j\right)\operatorname{Tr}\left(G_{ij}^{EM}G_{ij}^{EM\dagger}\right)\\
&+\operatorname{Im}\left(\chi_H^i\right)\operatorname{Im}\left(\chi_E^j\right)\operatorname{Tr}\left(G_{ij}^{ME}G_{ij}^{ME\dagger}\right)\\
&+\operatorname{Im}\left(\chi_H^i\right)\operatorname{Im}\left(\chi_H^j\right)\operatorname{Tr}\left(G_{ij}^{MM}G_{ij}^{MM\dagger}\right)\Big]
\end{aligned} \tag{26}$$

and the radiative power exchanged between particle $i$ and $j$ in many particles system can be calculated from

$$P_{j\leftrightarrow i} = P_{j\to i} - P_{i\to j} = 3\int_0^{+\infty}\frac{d\omega}{2\pi}\big(\Theta(\omega,T_j)-\Theta(\omega,T_i)\big)\mathcal{T}_{i,j}(\omega) \tag{27}$$

The net exchanged RHT power between two nanoparticles clusters considering many-body interaction obtained from CEMD is calculated from

$$\varphi = \sum_{j=1}^{Ne}\sum_{i=1}^{Na}P_{j\leftrightarrow i} \tag{28}$$

where $Ne$ is the number of particles in emitting cluster, and $Na$ is the number of particles in absorbing cluster. A definition of thermal conductance ($G$) between the two nanoparticles clusters is

$$G = \lim_{\delta T\to 0}\frac{\varphi}{\delta T} \tag{29}$$

where $\delta T$ is the temperature difference between emitting cluster and absorbing cluster. When MBI is not considered, namely, the existence of all other particles does not change the 'system Green function', hence the system Green function is just the Green function in vacuum, and the transmission coefficient between particle $i$ and $j$ is calculated from

$$\mathcal{T}_{i,j}^0(\omega) = \frac{4}{3}k^4 \Big[ \mathrm{Im}\big(\chi_E^i\big)\mathrm{Im}\big(\chi_E^j\big)\mathrm{Tr}\big(G_{0,ij}^{EE}G_{0,ij}^{EE\dagger}\big)$$

$$+ \mathrm{Im}\big(\chi_E^i\big)\mathrm{Im}\big(\chi_H^j\big)\mathrm{Tr}\big(G_{0,ij}^{EM}G_{0,ij}^{EM\dagger}\big)$$

$$+ \mathrm{Im}\big(\chi_H^i\big)\mathrm{Im}\big(\chi_E^j\big)\mathrm{Tr}\big(G_{0,ij}^{ME}G_{0,ij}^{ME\dagger}\big) \qquad (30)$$

$$+ \mathrm{Im}\big(\chi_H^i\big)\mathrm{Im}\big(\chi_H^j\big)\mathrm{Tr}\big(G_{0,ij}^{MM}G_{0,ij}^{MM\dagger}\big) \Big]$$

Then by omitting the MBI, the RHT power exchanged between two particles ($P_{j \leftrightarrow i}^0$), the net exchanged RHT power between two clusters ($\varphi_0$), and the thermal conductance without MBI ($G_0$) can be calculated using Eqs. (27), (28) and (29) with $\mathcal{T}_{i,j}^0(\omega)$, respectively. Note that this definition of the RHT without MBI is consistent with the previous definition by Dong et al. [29], which directly calculates RHT between two particles.

The thermal conductance calculated considering only EE contribution $G_{EE}$, namely, only using the first term in transmission coefficient in Eq. (26), is the same as the approach of the original many body radiative heat transfer theory [26], denoted as EP approach in the following for comparison.

## III. RESULTS AND DISCUSSION

Radiative heat transfer between two Ag nanoparticles clusters is investigated at various fractal dimensions ($D_f$) and separating gaps ($d$). Thermal conductance is calculated at 300K for all cases. The total thermal conductance is integrated over an angular frequency range from $0.1 \times 10^{14}$ rad·s$^{-1}$ to $90 \times 10^{14}$ rad·s$^{-1}$. A proper frequency resolution has been used to integrate spectral thermal conductance to obtain an accurate thermal conductance using the composite Simpson numerical integration method.

### A. RHT mechanism between metallic nanoparticle clusters

Thermal conductance between two Ag nanoparticles clusters as a function of separating gap ($d$) is shown in Fig. 3. Both $G$ and $G_{EE}$ are shown and the lines of $1/d^6$, $1/d^4$ and $1/d^2$ are added as reference. As shown, when $d$ is small (less than 1 μm), thermal conductance between two Ag nanoparticles clusters increases with the $D_f$. While $d$ is large enough (larger than 1 μm), $D_f$ has little effect on the thermal conductance. When $d$ is small enough, thermal conductance between two Ag nanoparticles is even larger than that of two Ag nanoparticles clusters. The reason to this phenomenon is that the distance between nanoparticles in proximity has priority to the number of the emitting and absorbing nanoparticles in determining the near-field thermal conductance. The straight-line distance between the two nanoparticles is usually smaller than the closest distance between two nanoparticles from the emitting and absorbing clusters, when the $D_f$ is not too large. As the $D_f$ increases, the straight-line distance between the two nanoparticles in proximity from the emitting and absorbing clusters approaches to that of two nanoparticles, which results in that $G$ between clusters is larger than that of two nanoparticles. The number of particles in clusters begins to dominate the near-field thermal conductance as $D_f$ of clusters increases to 2.3 and 2.8. Thermal conductance between two nanoparticles decays as $1/d^6$ in near field for $d$ less than 1μm and decays as $1/d^2$ in the far field. However, thermal conductance between two nanoparticles clusters decays slower than $1/d^4$ in near field and decays as $1/d^2$ in the far field. This may be attributed to the MBI in the nanoparticle clusters.

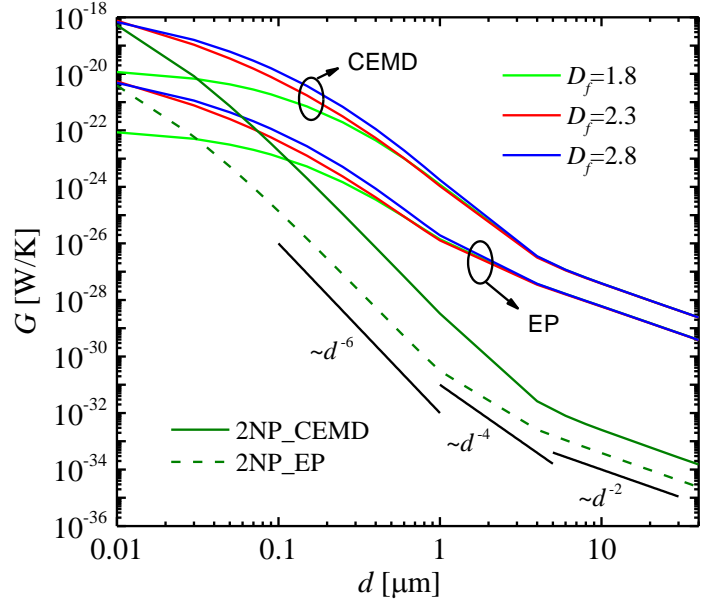

**Fig. 3** Thermal conductance between two Ag nanoparticles clusters at various fractal dimensions. Both CEMD approach and EP approach are used to calculate thermal conductance.

For dielectric nanoparticle, radiative heat transfer is dominated by the electric displacement current dissipation. While for metallic nanoparticle, eddy-current Joule dissipation due to a changing magnetic field in the particle dominates the radiative heat transfer. Thermal conductance obtained from the EP approach, considering only electric polarization response, is also shown in Fig. **3**. For both metallic nanoparticle clusters and two metallic nanoparticles, EP approach underestimates the radiative heat transfer as compare to the CEMD in both near field and far field.

To further explain the above observation on total thermal conductance, the spectral $G_\omega$ due to EE, EM, ME and MM contributions are presented in Fig. **4**, where $d$ is 1μm and $D_f$ is 2.8. For metallic nanoparticle clusters, MM contribution dominates the thermal conductance. $G_\omega$ due to EE contribution is far less than that of the MM contribution, which results in the underestimation of RHT using the EP approach. The peak of the $G_\omega$ due to MM contribution locates at 100 μm, which is consistent with the peak of imaginary part of magnetic polarizability of metallic Ag as shown in Fig. 2(b). However, the peaks of the $G_\omega$ due to EE, ME and EM contributions all locate at 10μm, which is corresponding with the characteristic thermal wavelength at 300 K.

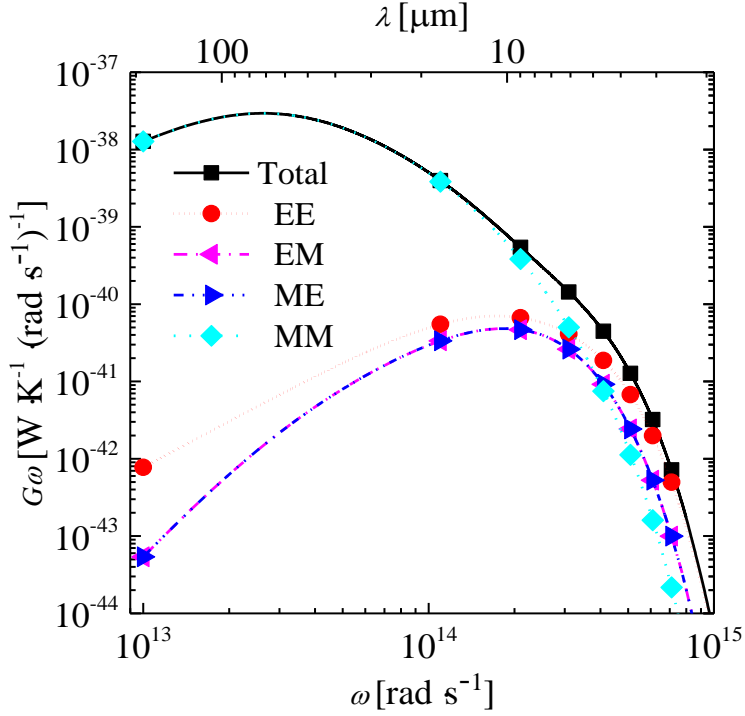

**Fig. 4** Spectral thermal conductance, $G_\omega$, due to EE, EM, ME and MM contribution, respectively. $D_f$ of the cluster is set as 2.8 and the separation gap between clusters is 1μm.

## B. Effect of many-body interaction

Previous studies reported that MBI inhibits the RHT in dielectric particle clusters [29, 38] though it enhances RHT in three SiC particle system [26]. It is still unclear whether MBI inhibits or enhances RHT in metallic nanoparticle clusters. In this section, MBI on RHT in metallic nanoparticle clusters is investigated. To evaluate the MBI on RHT, a definition of the enhancement factor of RHT due to MBI is given as

$$E = \frac{\varphi}{\varphi_0} \tag{31}$$

where $E$ is the enhancement factor, defined as the ratio of the net exchanged RHT power between two nanoparticles clusters considering MBI with that calculated without considering MBI. Both spectral $E$ and total $E$ can be easily calculated at a specified angular frequency and the angular frequency range of interest. In order to understand the MBI on RHT between metallic nanoparticle clusters, it is necessary to investigate the simplest case (two nanoparticles system) at first. The Enhancement factor of radiative thermal conductance between two Ag nanoparticles due to MBI is shown in Fig. 5. For two nanoparticles separated by 0.01μm, the spectral $E$ in the infrared frequency is nearly equal to 1, though in optical frequency range the $E$ is bigger than 1, which means that the MBI has little effect on the thermal radiative heat transfer between two Ag nanoparticles. Increasing separation gap between Ag particles decreases the enhancement in the optical frequency range due to MBI. From the point view of total

*E*, the MBI is unobvious for metallic nanoparticles, shown as the black line in Fig. 5(b). For dielectric nanoparticles, MBI inhibits slightly the RHT in the near field, while in the far field MBI has little effect on the RHT, shown as the red line in Fig. 5(b).

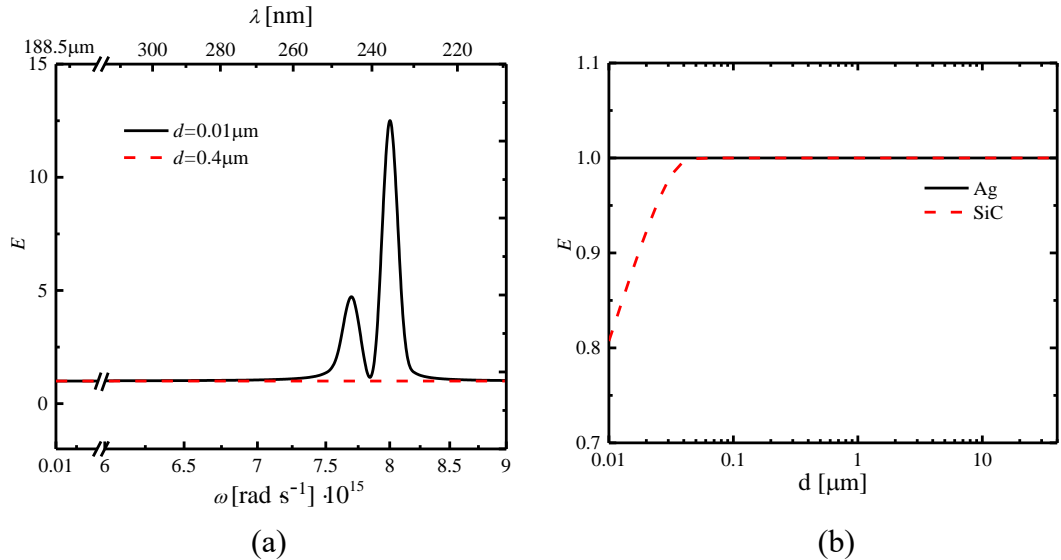

**Fig. 5** Enhancement factor of radiative thermal conductance between two Ag nanoparticles due to MBI: (a) spectral enhancement factor at two different separating gaps; (b) total enhancement factor as a function of separating gap, ranging from near field to far field.

The spectral enhancement factor between two Ag nanoparticles clusters at four different separating gaps is shown in Fig. 6(a). The spectral *E* increases dramatically in the optical frequency before resonance, which is corresponding with that between two nanoparticles. The spectral *E* in the infrared frequency is approaching to 1, which also means that the MBI has little effect on the thermal radiative heat transfer between Ag metallic nanoparticle clusters. Enhancement factor of RHT for both metallic nanoparticle clusters and dielectric clusters of SiC is shown as a function of *d* in Fig. 6(b). For dielectric nanoparticle clusters, MBI inhibits RHT in both near field and far field. In contrast, generally speaking, Enhancement factor of RHT for metallic nanoparticle clusters keeps constant (*E*=1) with various *d*, which means that MBI has unobvious effect on the RHT for metallic nanoparticle clusters in both near field and far field. Meanwhile, the inhibition of MBI on RHT for dielectric nanoparticle clusters decreases with the increasing separating gap, which is consistent with the results obtained by EP approach [29].

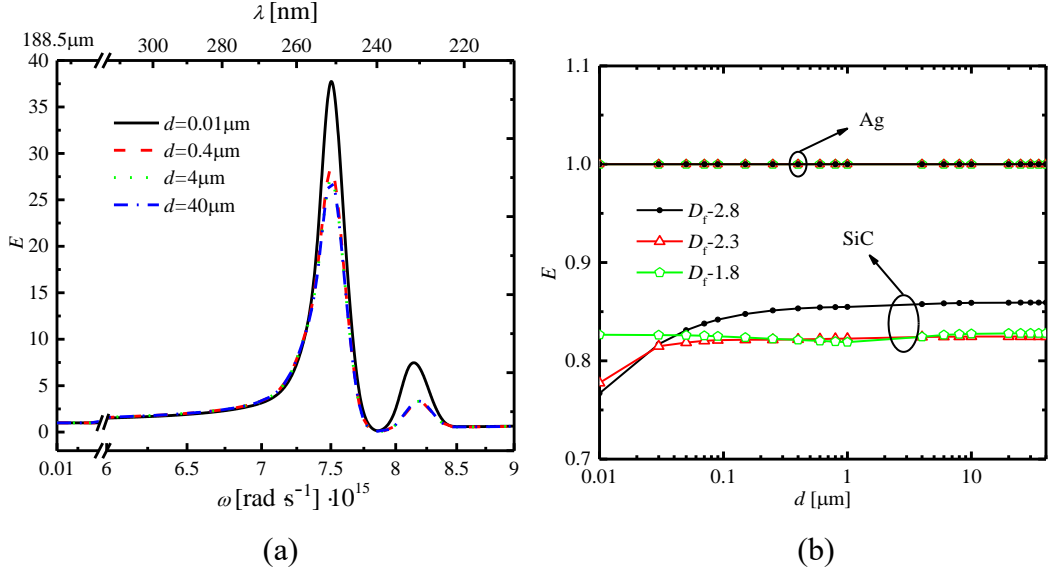

**Fig. 6** (a) Enhancement factor of RHT between two Ag nanoparticles clusters with $D_f$ 2.3 for different separation gaps. (b), enhancement factor of RHT for SiC and Ag nanoparticle clusters as a function of separating gap.

## C. Effect of relative orientation

Thermal conductance between two Ag nanoparticles clusters in three different relative orientations (parallel, oblique and vertical) is shown as a function of separation distance $d$ in Fig. **7**. When $d$ is larger than 3μm, both relative orientation and fractal dimension have no effect on the RHT. For clusters with high fractal dimension ($D_f$=2.8), rotation of the clusters has little effect on the RHT. When it comes to clusters with low fractal dimension ($D_f$=1.8), rotation of the clusters has significant effect on the RHT in near field. Nanoparticles from clusters in proximity and the number of nanoparticles from the emitting and the absorbing clusters play a dominant role in determining the NFRHT between nanoparticle clusters. For clusters of which the $D_f$ is 1.8, radiative heat flux of clusters with vertical orientation is much larger than that of the clusters with oblique and parallel orientation. The straight-line distance of particles in proximity from emitting and absorbing clusters with vertical orientation is much smaller than that of clusters with oblique and parallel orientation. In general, relative orientation has remarkable effect on radiative heat flux for clusters with lacy structure when the separation distance is in the near field. While for the separation distance in far field, both the relative orientation and the fractal dimension has a weak influence on radiative heat flux.

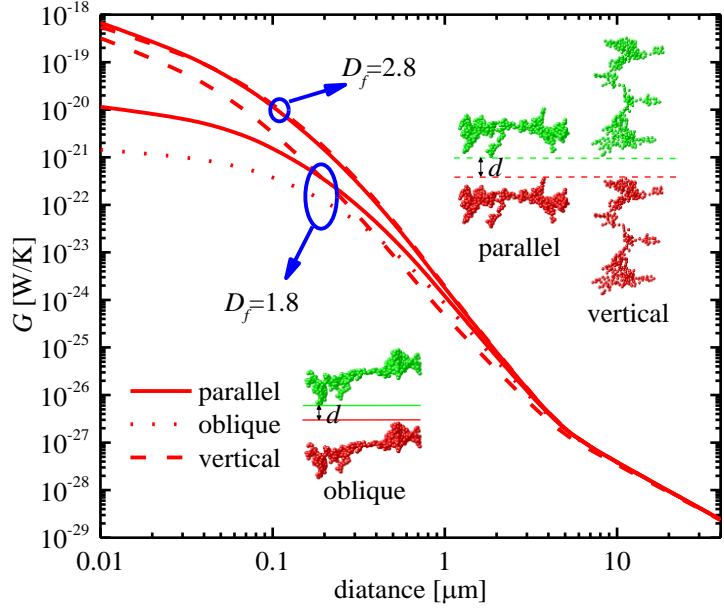

**Fig. 7** Thermal conductance between two Ag nanoparticles clusters in different orientations.

## IV. CONCLUSIONS

Near field effect is a key factor to influence thermal radiation transfer in dense particulate system when the particle separation distance is comparable to or less than the characteristic thermal wavelength. Near field radiative heat transfer between Ag metallic nanoparticle clusters in both near field and far field was studied by using the CEMD approach, considering contributions of all four electromagnetic field terms from electric and magnetic polarizations, namely EE, EM, ME and MM. The EP approach that considers only contribution of electric polarization underestimates the RHT in both near field and far field, which is attributed to the dominant role of MM contribution in the RHT between Ag metallic nanoparticle clusters. The effect of MBI on the RHT between Ag metallic nanoparticle clusters is unobvious at room temperature, while MBI inhibits the RHT between dielectric nanoparticle clusters. Effects of fractal dimension and relative orientation on RHT are also analyzed. When the separation distance is small (less than 1μm), thermal conductance between two Ag nanoparticles clusters increases with the fractal dimension. While when the separation distance is large enough, the fractal dimension shows little effect on the thermal conductance. The relative orientation has remarkable effect on radiative heat flux for clusters with lacy structure when the separation distance is in the near field. While for the separation distance in far field, both the relative orientation and the fractal dimension has a weak influence on radiative heat flux. A possible extension of this work is to take the interplay between the periodic configuration of many particles system and RHT into consideration [42] in the future.

## ACKNOWLEDGMENTS

The support by the National Natural Science Foundation of China (Grant nos. 51336002) and the Fundamental Research Funds for the Central Universities (Grant No. HIT.BRETIII.201415) was gratefully acknowledged.

## *Corresponding author*

[*] jmzhao@hit.edu.cn (Junming Zhao)